\documentclass{svproc}

\usepackage[a4paper,margin=2cm]{geometry}
\usepackage{graphicx, multirow, outlines, ulem}
\usepackage[utf8]{inputenc}
\usepackage{amsmath} 
\usepackage{amsfonts}
\usepackage{amssymb}
\usepackage{caption}
\usepackage{subcaption}
\usepackage{float}
\usepackage{cite}
\usepackage{epstopdf}
\DeclareGraphicsExtensions{.pdf,.png,.jpg}
\usepackage{breqn}
\usepackage{upgreek}

\usepackage{xcolor}
\colorlet{linkequation}{red}
\usepackage[colorlinks]{hyperref}

\usepackage{url}

\begin{document}
\mainmatter              
\title{Klein Tunneling in Uniaxial Strained Graphene under Super-Periodic Potential}
\titlerunning{Uniaxial Strained Graphene under SPP}  
%
\author{Sudhanshu Shekhar\inst{1} \and Bhabani Prasad Mandal\inst{2} \and Anirban Dutta\inst{3}}
\authorrunning{S. Shekhar et al.} 
%
%
\institute{$^{1,2,3}$ Department of Physics, Banaras Hindu University, Varanasi-221005, India\\
\email{ $^{1}$ sudhanshushekhar@bhu.ac.in $^{2}$ bhabani@bhu.ac.in $^{3}$ anirband@bhu.ac.in}}

\maketitle              

\begin{abstract}
In this article, we employ the transfer matrix method (TMM) to analytically explore the impact of uniaxial strain on electron scattering in graphene under locally periodic and super-periodic electrostatic potential. Our study reveals that strain significantly influences electron transmission through the merging parameter $(\delta)$, which modulates the Dirac cone positions. a positive merging parameter ($\delta > 0$) reduces the transmission probability by opening an energy gap at the merging point, while a negative merging parameter ($\delta < 0$) enhances transmission by bringing the Dirac cones closer, facilitating electron transport at certain angles.  However, the resonance peaks in super-periodic potential (SPP) are sharper for $\delta < 0$, making them more pronounced but increasingly difficult to resolve as the number of barriers increases.
\end{abstract}

\section{Introduction}

Graphene is a two-dimensional allotrope of carbon, arranged in a single-layer honeycomb lattice structure. Since its isolation in $2004$ \cite{novoselov2004electric}, it has garnered significant scientific interest due to its exceptional properties \cite{novoselov2004electric,castro2009electronic}. One of the most fascinating aspects of graphene's electronic properties is Klein tunneling \cite{nguyen2018klein, shekhar2024relativistic}, a phenomenon in which electrons can pass through potential barriers without backscattering. 

Beyond its properties, graphene's response to strain introduces new possibilities for material engineering. Unlike conventional materials, graphene can withstand strains exceeding $20\%$ without structural failure \cite{pereira2009tight}. However, when subjected to strong uniaxial strain exceeding $23\%$ \cite{pereira2009tight}, the deformation of its honeycomb lattice significantly alters its intrinsic electronic properties \cite{ guinea2012strain,  linnik2012effective,  pereira2009strain}. Such strain-induced modifications impact the band structure, leading to phenomena such as Dirac cone merging and the disappearance of Dirac points.

Graphene's exceptional electronic properties come from its unique energy structure near the Dirac points, where the conduction and valence bands meet. When uniaxial strain is applied, these properties change, influencing effects like Klein tunneling. Studying Klein tunneling in strained graphene is particularly interesting because strain alters the behavior of electrons, affecting their movement and interactions. To explore this further, we used the concept of SPP \cite{hasan_SPP} in graphene. An SPP is a locally periodic potential (LPP) that adds an extra layer of modulation to an already periodic structure \cite{mohammad2025polyadic, umar2023quantum, SINGH2023169236, narayan2023tunneling}. This additional periodicity can significantly impact the electronic properties of graphene.  By analyzing how SPP affects strained graphene, we aim to understand how this extra periodicity influences the transport behavior of electrons. These insights are valuable for advancing graphene-based electronic and quantum devices.

We structure this article as follows: In section~(\ref{sec:3}), we used the transfer matrix method (TMM) and obtained transmission probability for strained graphene under both LPP and SPP. We discuss the scattering behavior and the effects of uniaxial strain on the electron scattering behavior in graphene under LPP and SPP in section~(\ref{sec:4}).
\section{Klein tunneling in strained graphene through super-periodic rectangular potential barriers}\label{sec:3}
M. Hasan et al. introduced the concept of SPP in his paper \cite{hasan_SPP}, where he derived closed-form expressions for every element of the transfer matrix in the case of super periodic repetition of potentials. In this article, we used the similar formalism presented in paper \cite{shekhar2024relativistic}  for studying a special case ($y$-component of the momentum is conserved) of two-dimensional scattering of particles. In which, the expression of transmission probabilities of LPP and SPP of order-$2$ are given by:
\begin{eqnarray}
	T(N_{1})=\frac{1}{1+[|m_{12}|(U_{N_{1}-1}(\xi_{1}))]^{2}}\label{periodic}\,;\quad
	T(N_{1}, N_{2})=\frac{1}{1+[|m_{12}|(U_{N_{1}-1}(\xi_{1})U_{N_{2}-1}(\xi_{2}))]^{2}}\label{super-periodic}\,,
\end{eqnarray}
where $N_{1},~N_{2}$ are the number of potential barriers for LPP and SPP respectively. The terms $U_{N_{1}-1},~ U_{N_{2}-1}$ denote Chebyshev Polynomials (CPs) of second kind, $m_{12}$ is element of transfer matrix and $\xi_{1},~\xi_{2}$ are the arguments of CPs.
\subsection{Transfer matrix  for strained graphene}
When graphene is subjected to a strong uniaxial strain, one of the nearest-neighbor hopping amplitudes is significantly modified due to lattice deformation. However, for simplicity, we assume that the distance between carbon atoms remains unchanged. 
In this case, the electronic properties of graphene can be described using a two-dimensional lattice model consisting of two sublattices. The corresponding Hamiltonian near the merging points, known as the universal Hamiltonian, is obtained, which provides a deeper understanding of the low-energy behavior of quasiparticles in strained graphene \cite{montambaux2009universal}. The general form of the two-dimensional low-energy universal Hamiltonian, excluding the tilt of Dirac cones, is given by $(\hbar=1)$ \cite{montambaux2009universal, milicevic2019type}:
\begin{equation}
	H_{\text{universal}}= \left(\delta+\frac{q_{x}^{2}}{2m^{*}}\right)\sigma_{x}+c_{y}q_{y}\sigma_{y}\,,
\end{equation}
where $a$ is denoting the interatomic distance, $\sigma_{x}$, $\sigma_{y}$ are the Pauli matrices and $t$, $t^{\prime}$ as shown in figure (\ref{figg}). The effective mass is given by $ m^{*} = \frac{8}{3(2t+t^{\prime})a^{2}} $, while the quasiparticle velocity is expressed as $ c_{y} = \frac{3}{2} a t^{\prime} $. Additionally, the parameter $ \delta = t^{\prime} - 2t $ is referred to as the ``merging parameter." Here, $ a \approx 0.14 $ nm. These three parameters are real-valued. In this model, when graphene is subjected to tensile or compressive strain along a specific direction, one of the nearest-neighbor hopping terms, $ t^{\prime} $, is modified, while the remaining two, $ t $, remain unchanged.
We have applied an electrostatic potential barrier  that has a rectangular shape and is infinite along the $y$ axis
\begin{equation}
    V(x)=\begin{cases}
        V_{0},~~-a< x< a\\0,~~\text{otherwise}
    \end{cases}\,.
\end{equation}
The motion of the electrons is described by the Dirac equation:
\begin{equation}
    \left(\left(\delta+\frac{q_{x}^{2}}{2m^{*}}\right)\sigma_{x}+c_{y}q_{y}\sigma_{y}+V(x)\right)\psi(x,y)=E\psi(x,y)\,.
    \label{dirac equation}
\end{equation}
In the equation (\ref{dirac equation}), $E$ is the energy. We assume that the incident electron wave propagates at an angle $\phi$ with respect to the $x$-axis. We then proceed to express the components of the Dirac spinor $\psi_{1}$ and $\psi_{2}$ for the Hamiltonian $H=\left(\delta+\frac{q_{x}^{2}}{2m^{*}}\right)\sigma_{x}+c_{y}q_{y}\sigma_{y}+V(x)I$ ($I$ is the unit matrix) in a subsequent manner:
\begin{eqnarray}
    \psi_{1}(x,y)=\begin{cases}
        (Ae^{ik_{x}x}+Be^{-ik_{x}x})\psi(y),x<-a\\(Ee^{iq_{x}x}+Fe^{-iq_{x}x})\psi(y),-a<x<a\\(Ce^{ik_{x}x}+De^{-ik_{x}x})\psi(y),x>a
    \end{cases}\,;
    \label{}     \psi_{2}(x,y)=\begin{cases}
        s(Ae^{ik_{x}x+i\phi}-Be^{-ik_{x}x-i\phi})\psi(y),x<-a\\s'(Ee^{iq_{x}x+i\theta}-Fe^{-iq_{x}x-i\theta})\psi(y),-a<x<a\\s(Ce^{ik_{x}x+i\phi}-De^{-ik_{x}x-i\phi})\psi(y),x>a
    \end{cases}\,,
    \label{}
\end{eqnarray}
where $k_{x}=E\cos{\phi}$, $k_{y}=E\sin{\phi}$ are $x$ and $y$-components of the wavevector outside of the barrier respectively, $\psi(y)=e^{ik_{y}y}$, $q_{x}=\sqrt{(E-V_{0})^{2}-k_{y}^{2}c_{y}^{2}}$, $\theta=\tan^{-1}\left(\frac{k_{y}}{q_{x}}\right)$, $s=\text{sgn}(E)$, and $s'=\text{sgn}(E-V_{0})$.
Using the continuity of the wave function at the boundaries $x=-a$ and $x=a$, one can find the transfer matrix, and the elements of the transfer matrix are:
\begin{subequations}
    \begin{align}
    m_{11} & = \frac{1}{ss'}e^{2ik_{x}a} (ss'\cos (2q_{x}a)-i\sin(2q_{x}a)(\sec(\theta)\sec(\phi)-ss'\tan(\theta )\tan (\phi)))=m_{22}^{*}\label{m11}\,,\\m_{12} & = \frac{1}{2ss'}\sec(\theta)e^{-i(2k_{x}a+\phi)}\sin(2q_{x}a)(-2ss'\sin(\theta)\sec(\phi)+2\tan(\phi))=m_{21}^{*}\label{m12}\,.
    \end{align}
\end{subequations}
\subsection{Tunneling through LPP and SPP of order-2}
The transfer matrix can be used to calculate the transmission probability for the SPP of any order $N_{n}$. This can be done using equation (\ref{periodic}), with equation (\ref{m12}). For completeness, we provide the expression for the transmission probability $T(N_{1},\phi)$ for the periodic barrier or super periodic barrier of order-1;
\begin{equation}
    T(N_{1},\phi)=\frac{1}{1+[\sin(2q_{x}a) (s s' \tan (\theta ) \sec (\phi )-\sec (\theta ) \tan (\phi ))U_{N_{1}-1}(\xi_{1})]^{2}}\label{periodic transmission}\,.
\end{equation}
Next, the transmission probability $T(N_{1},N_{2},\phi)$ for super periodic electrostatic barrier of order-$2$,
\begin{equation}
	T(N_{1},N_{2},\phi)=\frac{1}{1+[\sin(2q_{x}a) (s s' \tan (\theta ) \sec (\phi )-\sec (\theta ) \tan (\phi ))U_{N_{1}-1}(\xi_{1})U_{N_{2}-1}(\xi_{2})]^{2}}\label{spp}\,.
\end{equation}

The expressions of transmission probabilities for normal graphene in the LPP and SPP of order-$2$ are given by \cite{shekhar2024relativistic}:
\begin{eqnarray}
	T(N_{1},\phi)_{\text{normal}}=\frac{1}{1+[\sin(2q_{x}a) (s s' \tan (\theta ) \sec (\phi )-\sec (\theta ) \tan (\phi ))U_{N_{1}-1}(\xi_{1})]^{2}}\,;\\
	T(N_{1},N_{2},\phi)_{\text{normal}}=\frac{1}{1+[\sin(2q_{x}a) (s s' \tan (\theta ) \sec (\phi )-\sec (\theta ) \tan (\phi ))U_{N_{1}-1}(\xi_{1})U_{N_{2}-1}(\xi_{2})]^{2}}\,.
\end{eqnarray}
The expressions for transmission probabilities under uniaxial strain in LPP and SPP have the same mathematical form as those for normal graphene, with only the parameter definitions being different.
\section{Result and discussion}\label{sec:4}
\begin{figure}[htb]
\vspace{-0.5cm}
	\centering
	\begin{subfigure}[b]{0.23\textwidth}
		\centering
		\includegraphics[height=4cm]{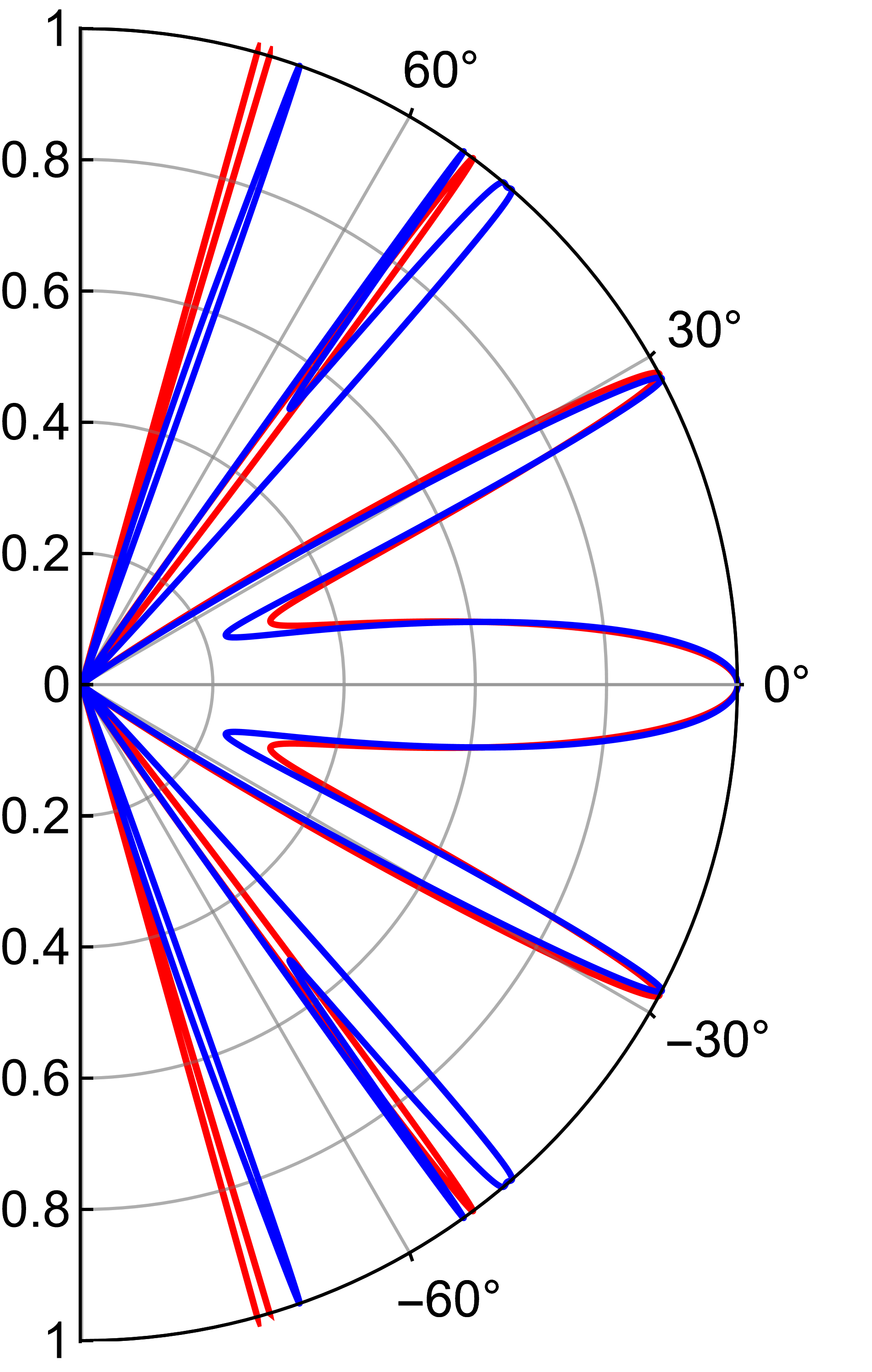}
		\caption{$N_{1}=3$}
		\label{1a}
	\end{subfigure}
	\begin{subfigure}[b]{0.23\textwidth}
		\centering
		\includegraphics[height=4cm]{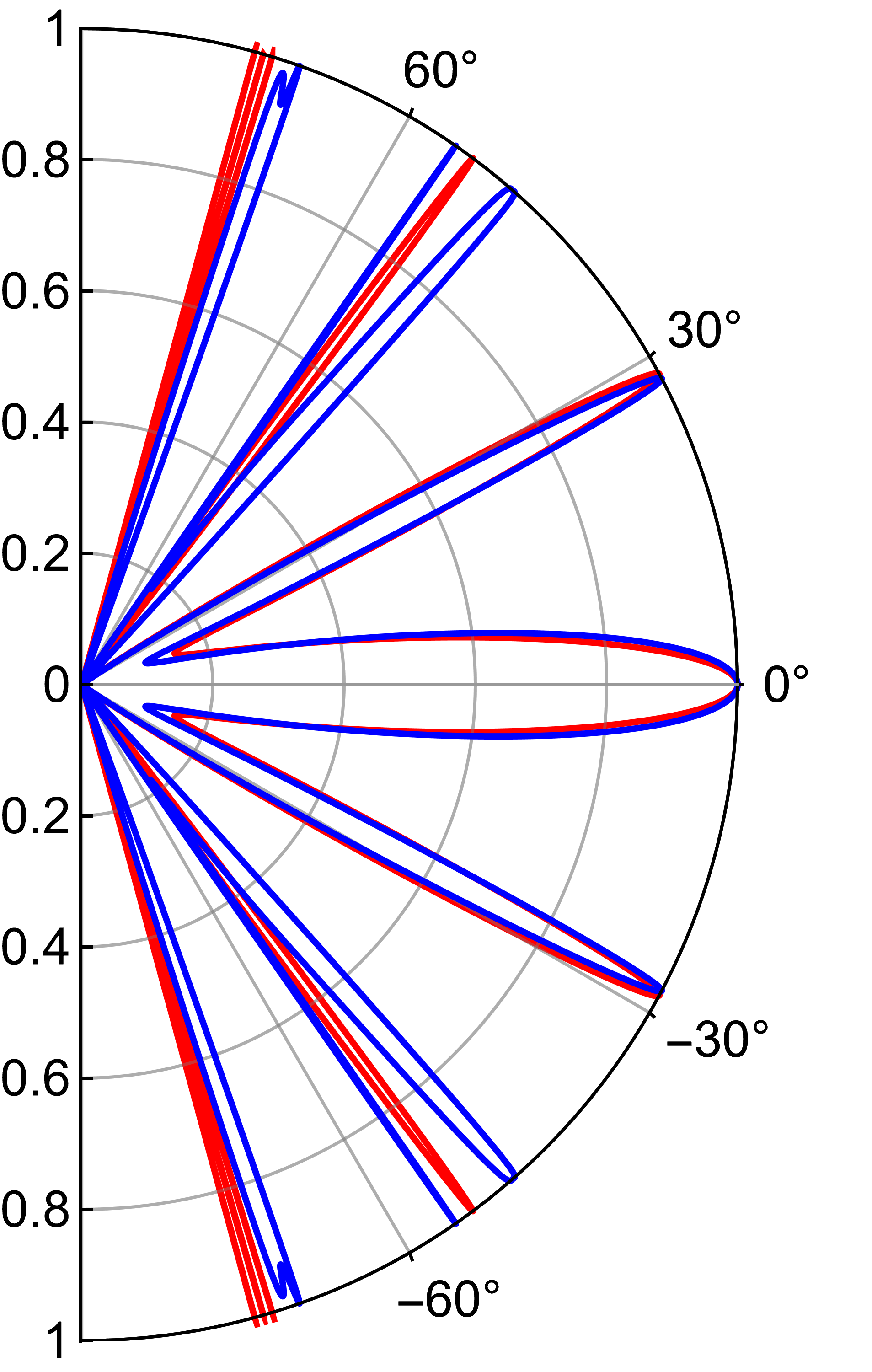}
		\caption{$N_{1}=4$}
		\label{1b}
	\end{subfigure}
	\begin{subfigure}[b]{0.23\textwidth}
		\centering
		\includegraphics[height=4cm]{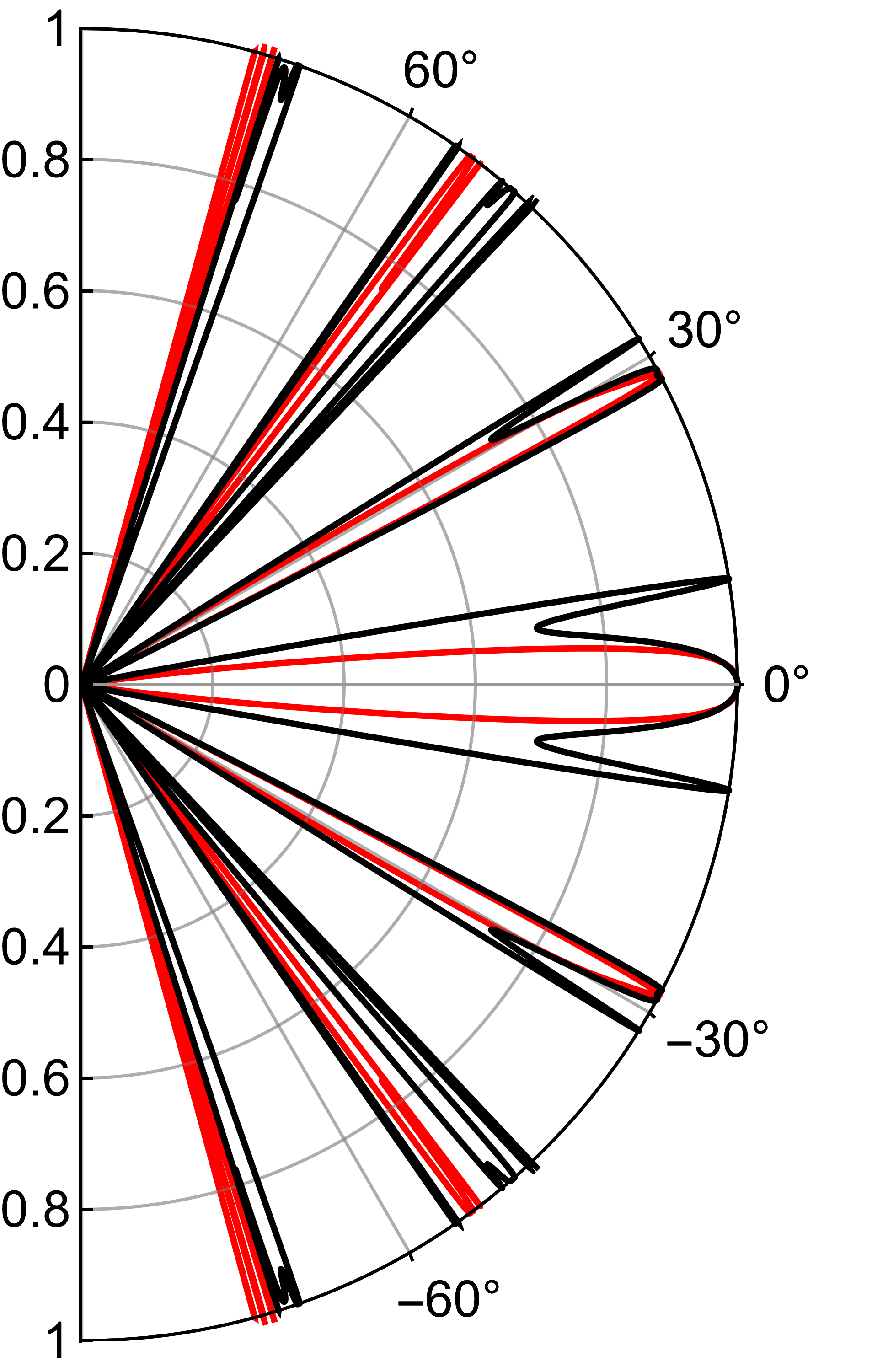}
		\caption{$N_{1}=4,~N_{2}=3$}
		\label{1c}
	\end{subfigure}
	\begin{subfigure}[b]{0.23\textwidth}
		\centering
		\includegraphics[height=4cm]{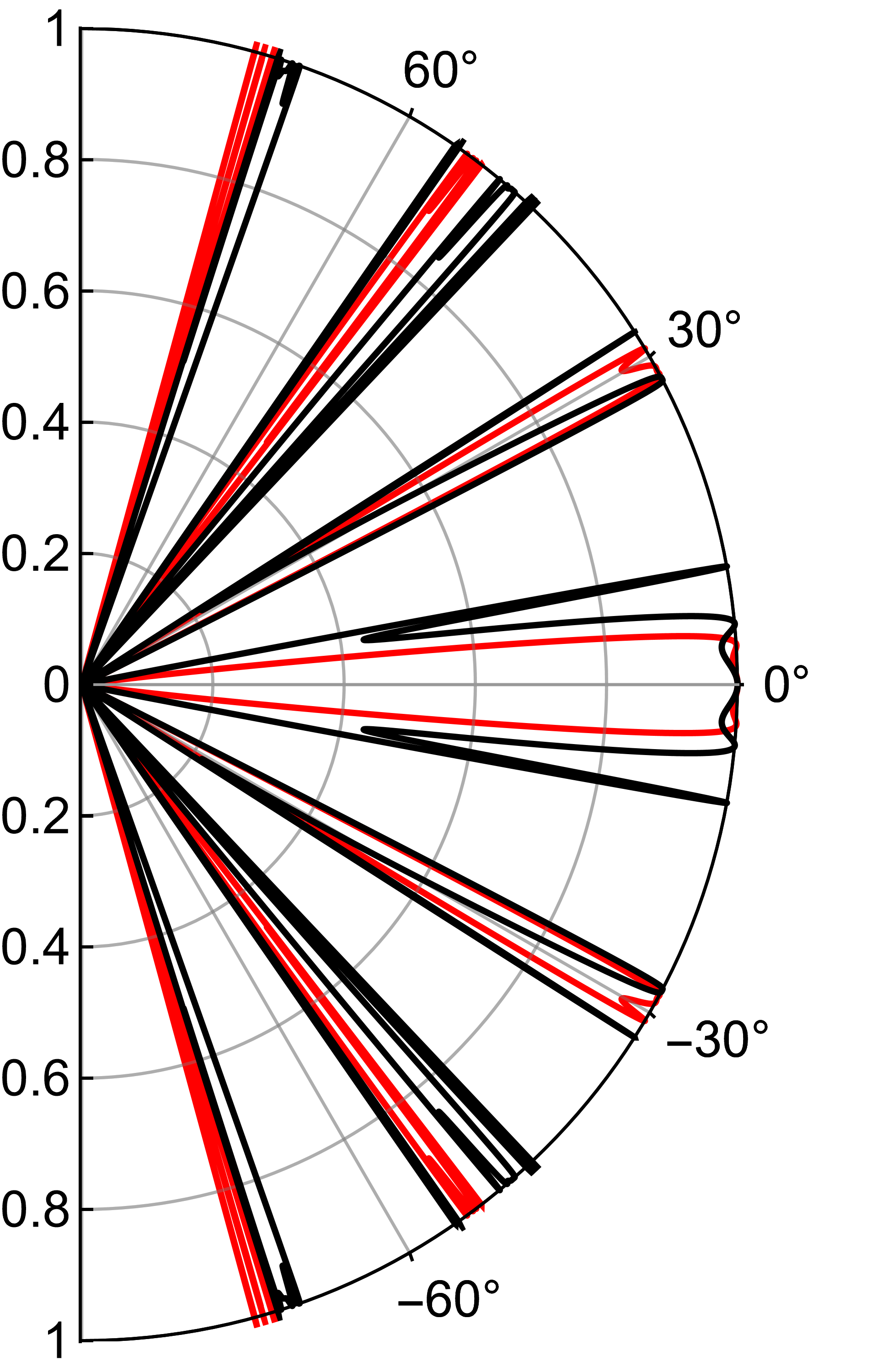}
		\caption{$N_{1}=4,~N_{2}=4$}
		\label{1d}
	\end{subfigure}
	\begin{subfigure}[b]{0.23\textwidth}
		\centering
		\includegraphics[height=4cm]{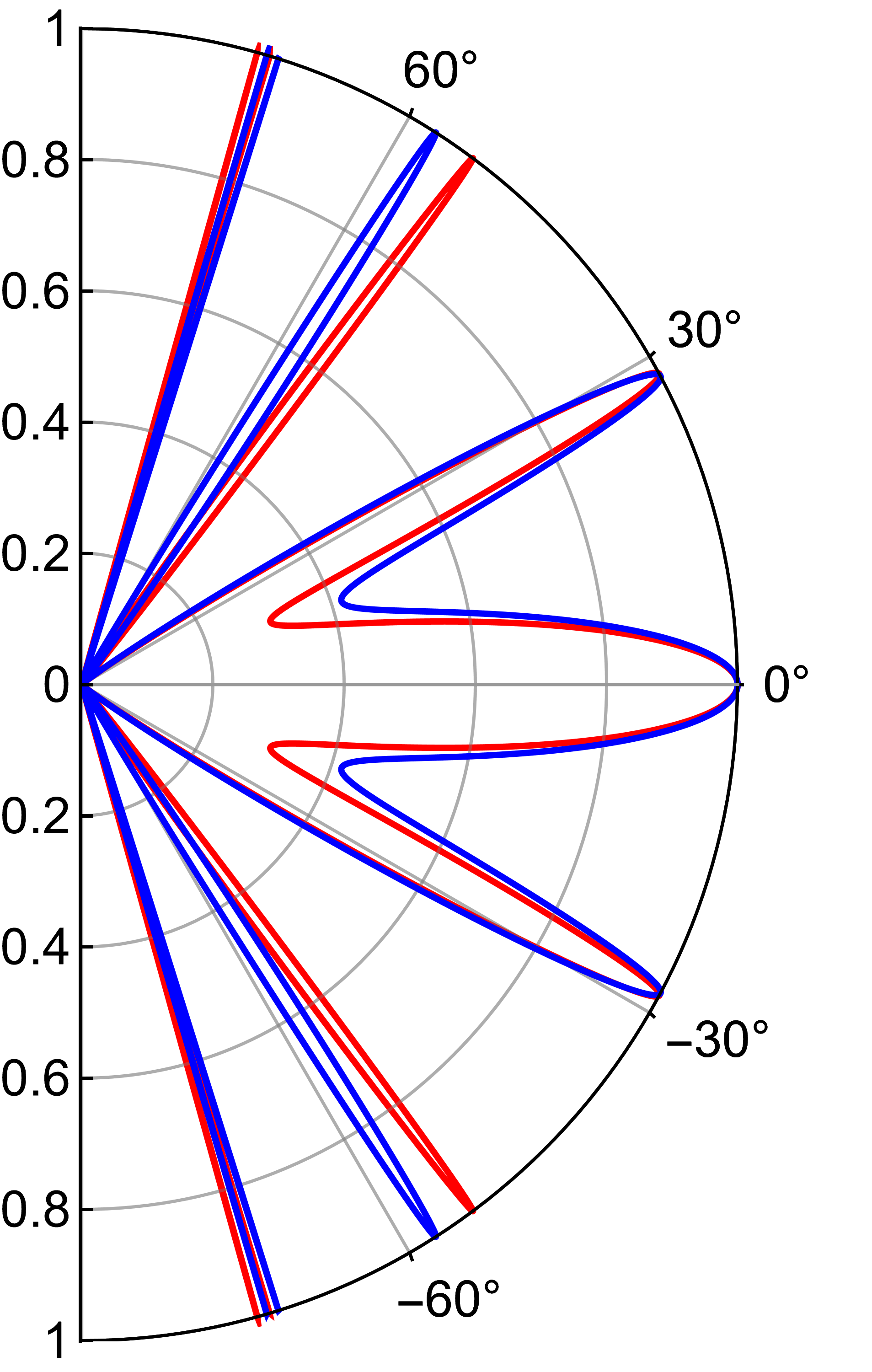}
		\caption{$N_{1}=3$}
		\label{1e}
	\end{subfigure}
	\begin{subfigure}[b]{0.23\textwidth}
		\centering
		\includegraphics[height=4cm]{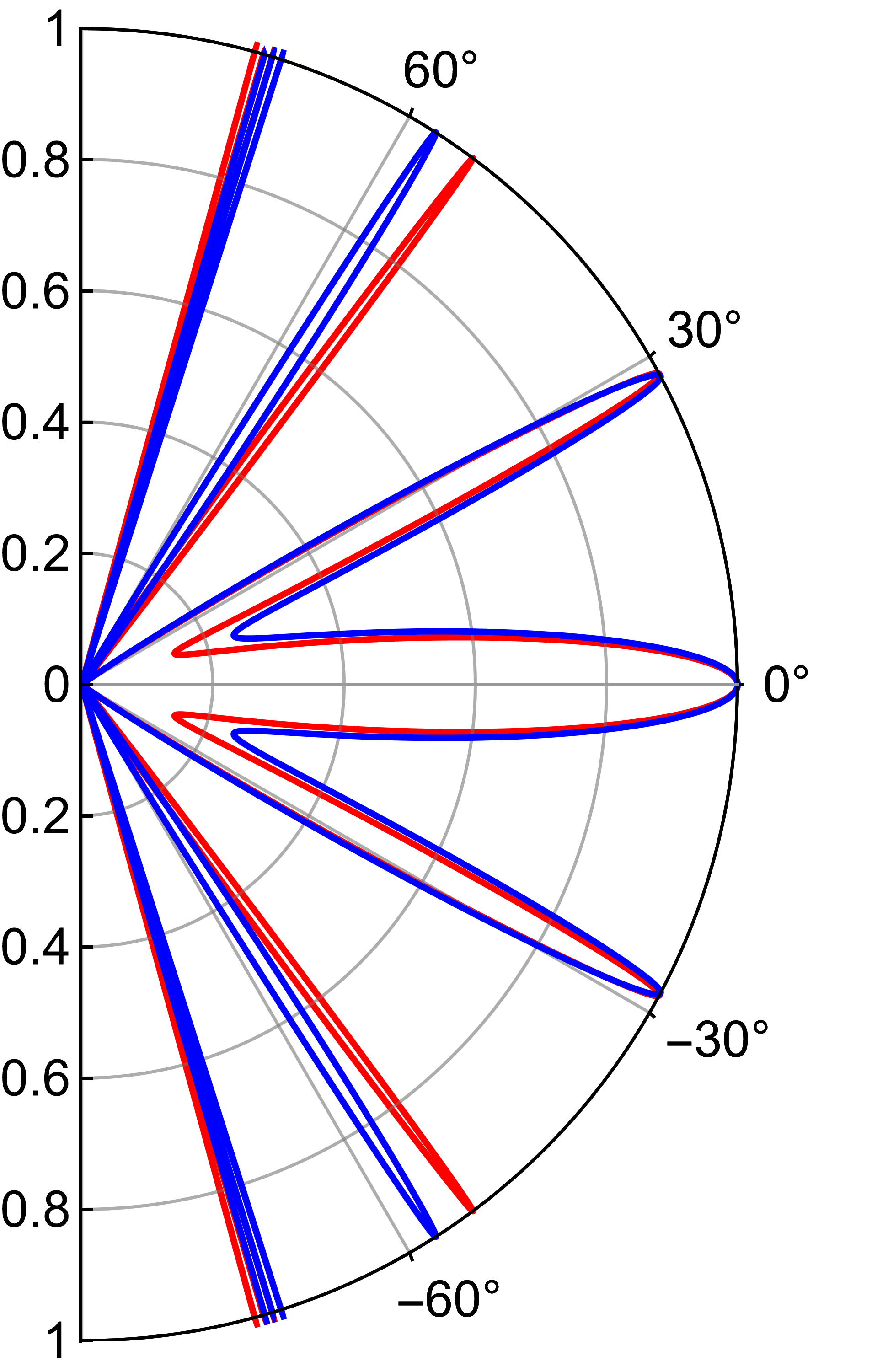}
		\caption{$N_{1}=4$}
		\label{1f}
	\end{subfigure}
	\begin{subfigure}[b]{0.23\textwidth}
		\centering
		\includegraphics[height=4cm]{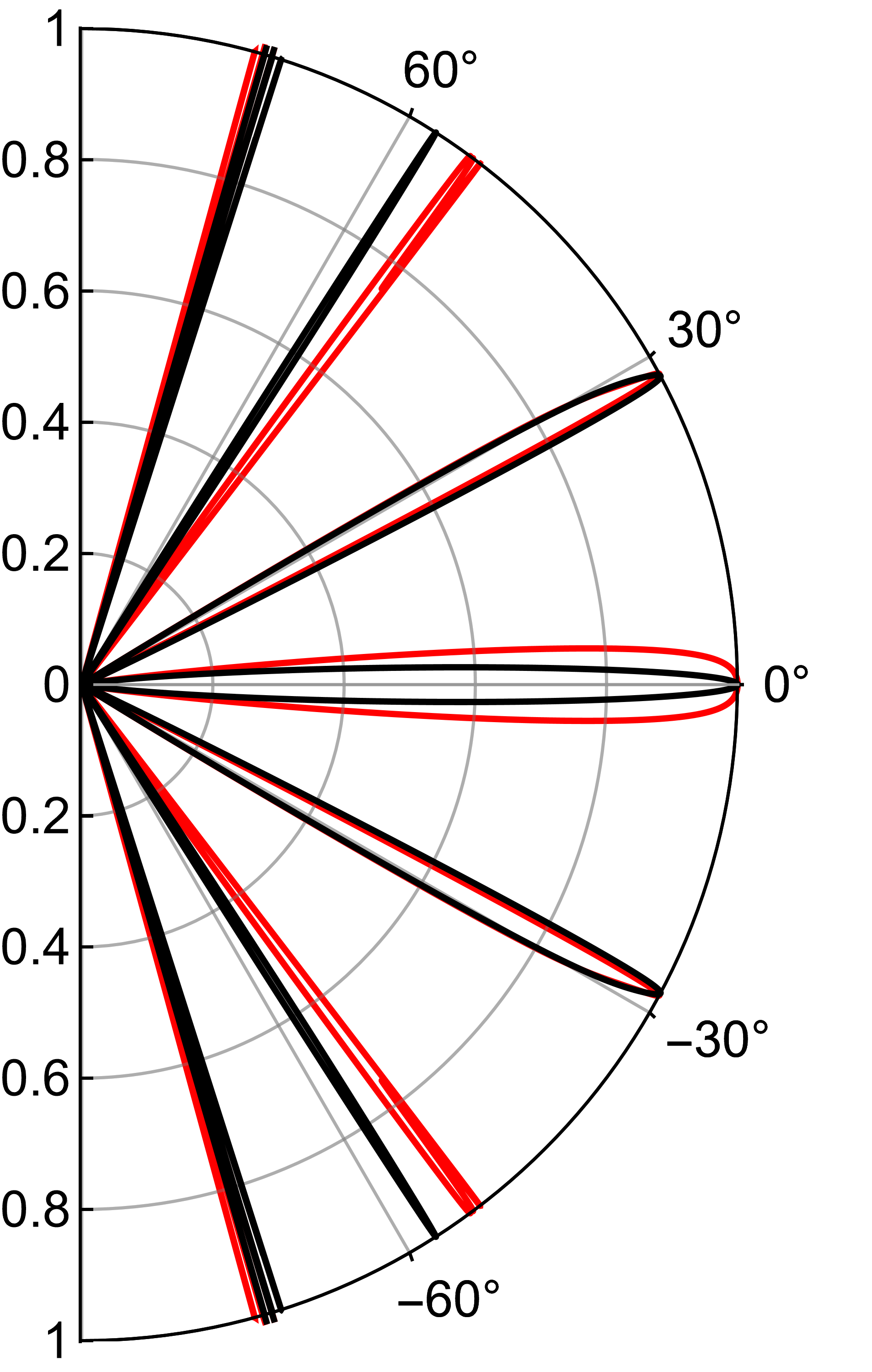}
		\caption{$N_{1}=4,~N_{2}=3$}
		\label{1g}
	\end{subfigure}
	\begin{subfigure}[b]{0.23\textwidth}
		\centering
		\includegraphics[height=4cm]{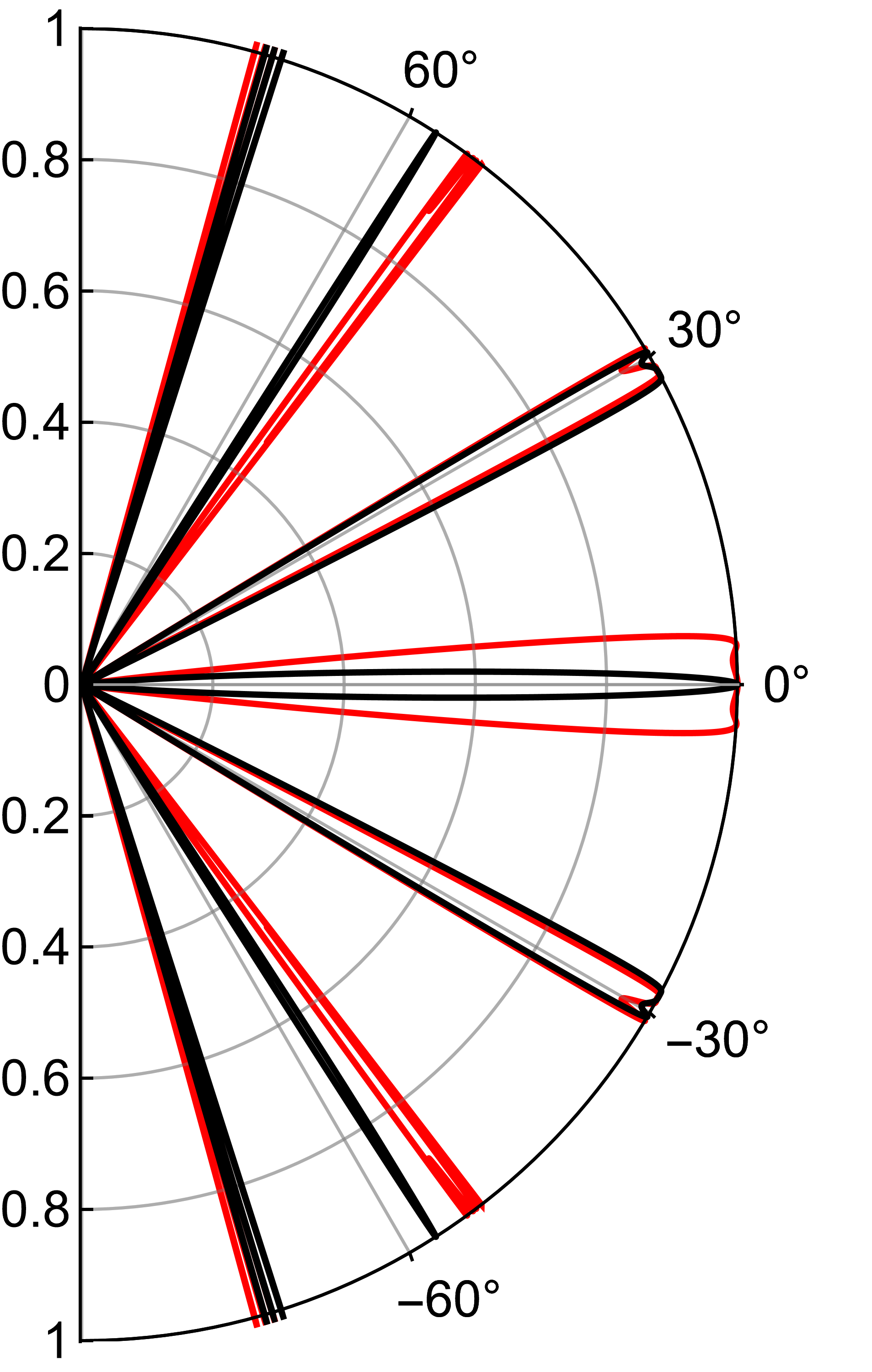}
		\caption{$N_{1}=4,~N_{2}=4$}
		\label{1h}
	\end{subfigure}
	\caption{Polar plot of transmission probability for normal (red) and uniaxial strained graphene (blue) for LPP and SPP of order-$2$.}
	\label{fig: N1}
    \vspace{-0.7cm}
\end{figure}

The scattering behavior of electrons for normal graphene in periodic and SPP has been studied in \cite{shekhar2024relativistic}. Here, we explore how uniaxial strain affects this behavior in the presence of such potentials.\\
Equations (\ref{periodic transmission}) and (\ref{spp}) show that perfect transmission occurs at $\phi=0$, $2q_{x}a=\beta\pi$ with $\beta$ an integer and also the CP is zero.
\\
Figures (\ref{1a}),~(\ref{1b}),~(\ref{1e}) and figure~(\ref{1f}) illustrate the transmission probability for a LPP as a function of the incident angle at a fixed energy of  $E = 80$ meV. The number of barriers varies from $N_{1} = 3 $ to $N_{1} = 5$. In this case, the barrier width and the spacing between successive barriers are $2a = 200$ nm and $c_{1} = 90$ nm, respectively. The merging parameter is $\delta = 20$ meV for figures (\ref{1a}),~(\ref{1b}),~(\ref{1c}) and figure~(\ref{1d}) and $\delta = -20$ meV for figures (\ref{1e}),~(\ref{1f}),~(\ref{1g}) and figure~(\ref{1h}). The barrier height is fixed at $V_{0} = 220$ meV.
\\
In this figure~(\ref{fig: N1}), it can be seen that at $\phi=0^\circ$, the transmission coefficient equals unity and does not depend on the number of barriers for normal and strained graphene. This behavior confirms the Klein-tunneling effect in LPP and SPP, which states that the system is completely transparent for normal incidence, even for large barrier widths. 
\begin{figure}[htb]
\vspace{-0.7cm}
    \begin{minipage}{0.40\textwidth}
        \centering
        \includegraphics[height=3cm]{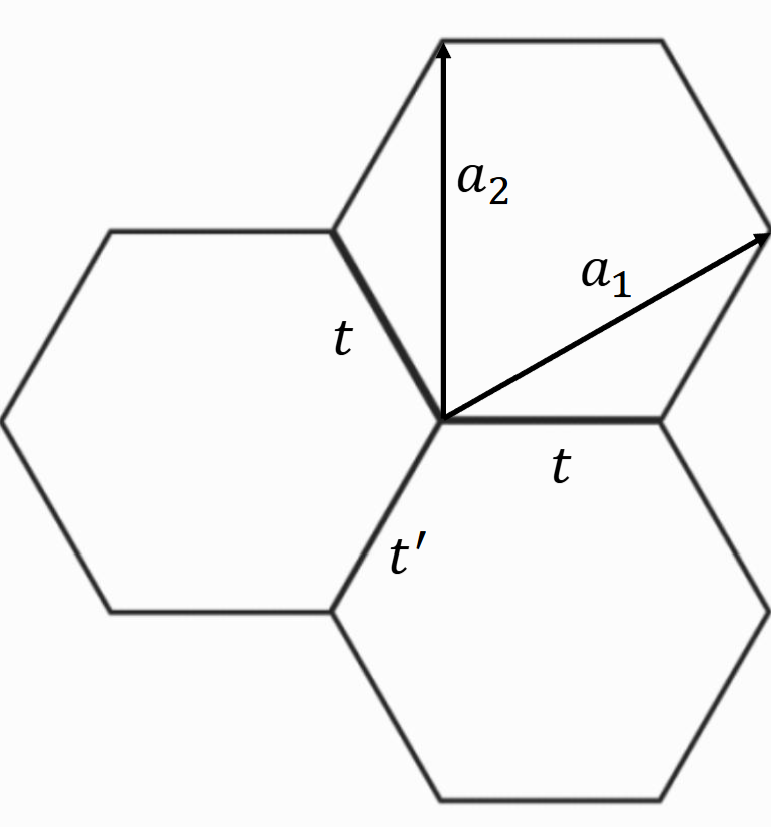}
        \caption{The $t-t^{\prime}$ model. The Bravais lattice vectors $\vec{a_{1}}=\frac{\sqrt{3}a}{2}\hat{i}+\frac{3a}{2}\hat{j},~ \vec{a_{2}}=\frac{\sqrt{3}a}{2}\hat{i}+\frac{3a}{2}\hat{j}$ and the hopping amplitudes $t$ and $t^{\prime}$ are indicated.}
        \label{figg}
    \end{minipage}%
    \hspace{1cm} 
    \begin{minipage}{0.40\textwidth}
        \vspace{0.3cm} 
        \begin{tabular}{ |p{2.1cm}|p{2.1cm}|p{2.1cm}|p{2.1cm}|  }
            \hline
            No. of barrier $(N_{1})$ &$\lim_{\phi\to\frac{\pi}{9}}T_{\text{normal}}$ & $\lim_{\phi\to\frac{\pi}{9}}T$ at $\delta = 20 \text{ meV}$ &$\lim_{\phi\to\frac{\pi}{9}}T$ at $\delta = -20 \text{ meV}$ \\
            \hline
            $N_{1} = 2$ & $0.5683$ & $0.4870$ &$0.6715$ \\
            $N_{1} = 3$ & $ 0.3147$& $0.2385$   & $0.4386$ \\
            $N_{1} = 4$ &$ 0.1576$&$0.1064$ & $0.2638$ \\
            $N_{1} = 5$  & $ 0.0751$  &$ 0.0456$ & $0.1514$ \\
            \hline
        \end{tabular}
        \captionof{table}{Transmission probability in normal and strained graphene.}
        \label{table1}
    \end{minipage}
\vspace{-0.8cm}
\end{figure}

As the number of barriers increases at a specific incident angle, the transmission probability decreases in both normal and strained graphene, as shown in Table~(\ref{table1}). For a positive merging parameter, the transmission probability is lower than in normal graphene. This occurs because a positive merging parameter induces an energy gap at the merging point, leading to reflection at certain incident angles. Conversely, for a negative merging parameter, the transmission probability is higher than in normal graphene. This is because a negative merging parameter brings the Dirac cones closer at the merging point, creating a pathway for electron transmission through the system at specific incident angles.
\\
Now, We will explore the transmission probability for the super-periodic electrostatic potential of order-$2$ in strained graphene. Figures (\ref{1c}),~(\ref{1d}),~(\ref{1g}) and figure~(\ref{1h}) show the transmission probability for the super periodic electrostatic potential of order -$2$ as a function of the incident angle for a fixed energy of $E = 80\text{ meV}$. The number of barriers of SPP are $N_{2} = 3$ and $N_{2} = 4$ and the fix value of $N_{1}=4$. In this case, the separation between successive barriers is $c_{2} = 60\text{ nm}$, and other parameters are the same as the figure (\ref{1a}).
\\
The overall transmission probability for both positive and negative values of the merging parameter for SPP follows a trend similar to that observed in LPP. However, for $ \delta < 0 $, the resonance peaks \cite{shekhar2024relativistic, hasan_SPP} are sharper than those for $ \delta > 0 $, as shown in figures (\ref{1c}),~(\ref{1d}),~(\ref{1g}) and figure~(\ref{1h}). Additionally, resolving the resonance peaks becomes increasingly difficult for higher $ N_{2} $ when $ \delta < 0 $.
\begin{figure}[htb]
\vspace{-0.5cm}
	\centering
	\begin{subfigure}[b]{0.23\textwidth}
		\centering
		\includegraphics[height=4cm]{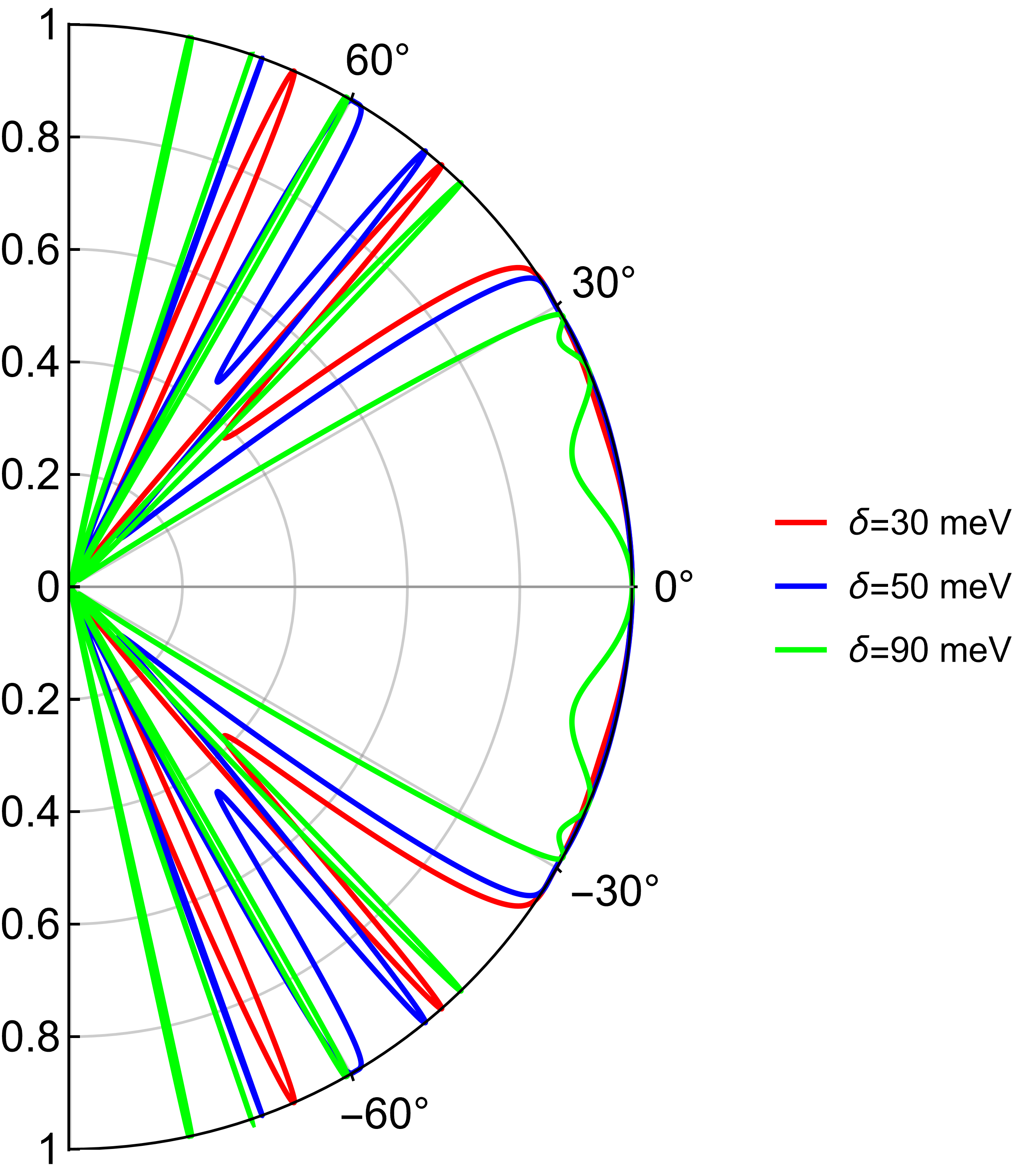}
		\caption{$N_{1}=2$}
		\label{fig: deltaa}
	\end{subfigure}
	\begin{subfigure}[b]{0.23\textwidth}
		\centering
		\includegraphics[height=4cm]{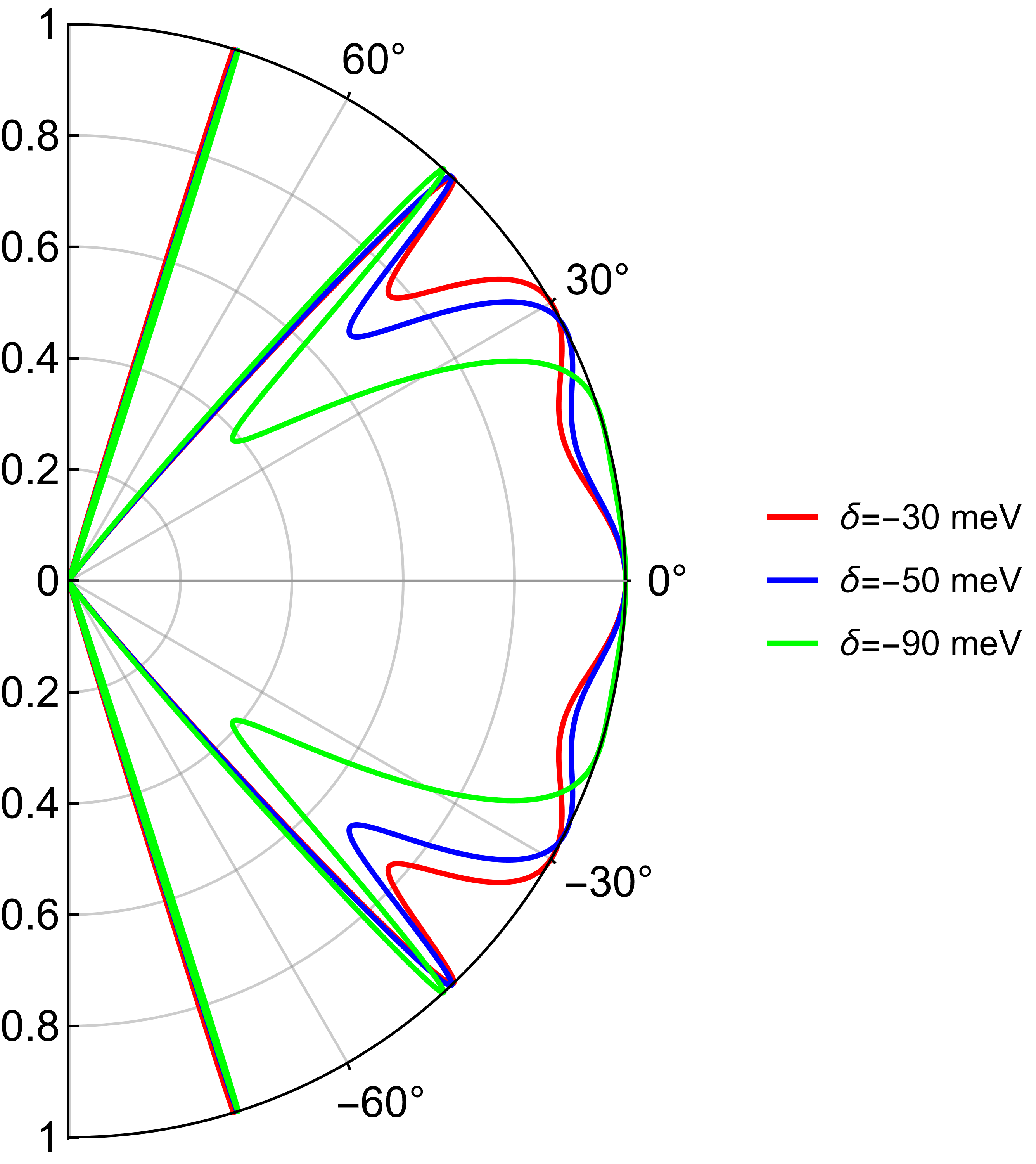}
		\caption{$N_{1}=2$}
		\label{fig: deltab}
	\end{subfigure}
	\begin{subfigure}[b]{0.23\textwidth}
		\centering
		\includegraphics[height=4cm]{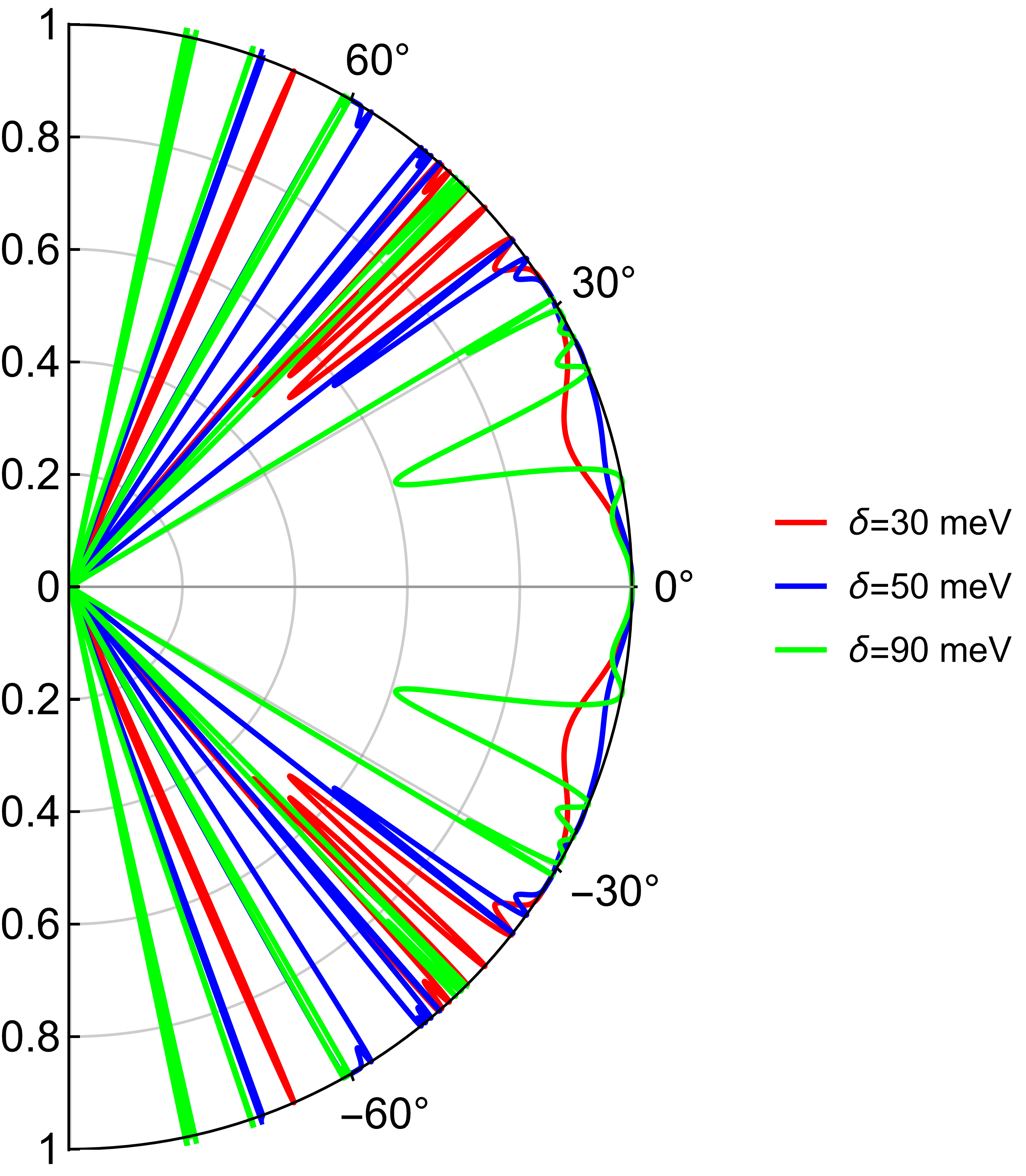}
		\caption{$N_{1}=2,~N_{2}=4$}
		\label{fig: deltac}
	\end{subfigure}
	\begin{subfigure}[b]{0.23\textwidth}
		\centering
		\includegraphics[height=4cm]{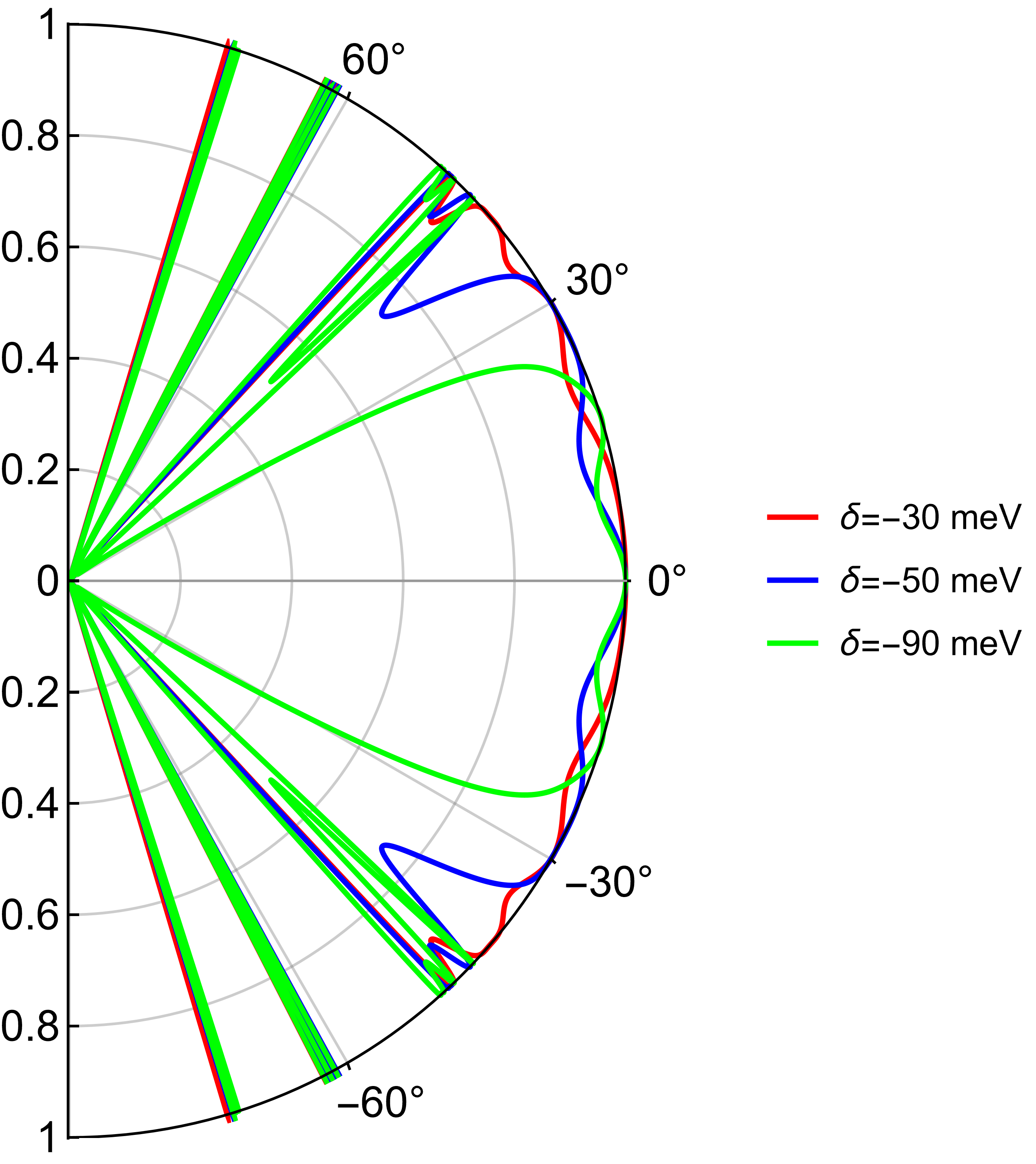}
		\caption{$N_{1}=2,~N_{2}=4$}
		\label{fig: deltad}
	\end{subfigure}
	\caption{Polar plot of transmission probability at different values of $\delta$ for both LPP (a, b) and SPP of order-$2$ (c, d). In these plots, the barrier width and the spacing between successive barriers are $2a = 200$ nm, and $c_{1} = 90$ nm, for LPP, while for the SPP of order-$2$, the spacing is $c_{2} = 60$ nm. The energy and the barrier height are fixed at $E=70$ meV and $V_{0} = 240$ meV respectively.}
	\label{fig: Nd}
    \vspace{-0.7cm}
\end{figure}
\\
At certain lower incident angles, the transmission probability is relatively high for smaller positive $\delta$ values in both locally periodic and SPP of order-2, as shown in figures~(\ref{fig: deltaa}) and (\ref{fig: deltac}). Conversely, for larger negative $\delta$ values, the transmission probability is relatively low in both cases, as depicted in figures~(\ref{fig: deltab}) and (\ref{fig: deltad}). In general, as the magnitude of $\delta$ increases, the transmission probability decreases, making the barrier more opaque for larger negative $\delta$ values and less opaque for larger positive $\delta$ values.

\end{document}